# Observation of Strong Nonreciprocal Thermal Emission

Zhenong Zhang[1†], Alireza Kalantari Dehaghi[1†], Pramit Ghosh[1], Linxiao Zhu[1*]

[1]Department of Mechanical Engineering, The Pennsylvania State University, University Park, PA 16802, United States.

†These authors contributed equally to this work.

*Corresponding author. Email: lqz5242@psu.edu

**Abstract**

The Kirchhoff's law of thermal radiation stating the equivalence of emissivity and absorptivity at the same wavelength, angle, and polarization, has completely constrained emission and absorption processes. Achieving strong nonreciprocal emission points to fundamental advances for applications such as energy harvesting, heat transfer, and sensing, but strong nonreciprocal thermal emission has not been experimentally realized. Here, we observe strong nonreciprocal thermal emission using a custom-designed angle-resolved magnetic thermal emission spectroscopy and an epitaxially-transferred gradient-doped metamaterial. We show that under magnetic field, the metamaterial strongly breaks the Kirchhoff's law, with a difference between emissivity and absorptivity at the same wavelength and angle reaching as high as 0.43. Significant nonreciprocal emission persists over broad spectral and angular ranges. The demonstration of strong nonreciprocal thermal emission and the approach can be useful for systematic exploration of nonreciprocal thermal photonics for thermal management, infrared camouflage, and energy conversion.

The Kirchhoff's law of thermal radiation states that, for an arbitrary object, the emissivity must equal the absorptivity, at the same wavelength, angle, and polarization [1-3]. Such reciprocal relation between emission and absorption has played a foundational role in understanding and controlling emission [4-8]. Breaking the reciprocal relation between emission and absorption can not only provide separate control of emission and absorption, but also point to energy harvesting at thermodynamic limits such as in solar cells [9,10], thermophotovoltaics [11,12] and harvesting outgoing radiation [11,13], advanced heat flux control [14-17] and communication [18].

The Kirchhoff's law of thermal radiation is not required by thermodynamics, but rather is the consequence of Lorentz reciprocity [19,20]. It has been pointed out that nonreciprocal emission and absorption can be achieved by breaking Lorentz reciprocity [21-25], leading to unequal emissivity $e(\lambda, \theta)$ and absorptivity $\alpha(\lambda, \theta)$ with $e(\lambda, \theta) \neq \alpha(\lambda, \theta)$, as illustrated in Fig. 1(a). While there have been great advances in the theory of nonreciprocal emission and absorption in magneto-optical materials [26-32] and magnetic Weyl semimetals [33-35], experiment of nonreciprocal emission is underexplored and the achieved nonreciprocity in existing demonstration of nonreciprocal emission [36,37] defined as the difference between emissivity and absorptivity is small. Recent experiments measured nonreciprocal emission from heated InAs-based emitter under magnetic field [36,37], with a nonreciprocity ~0.22 and a narrow bandwidth in Ref. [36], and a nonreciprocity up to ~0.12 and a bandwidth of 3.5 μm in Ref. [37]. Though strong nonreciprocal emission is necessary for achieving energy conversion at thermodynamic limits and nonreciprocal heat flux control [28], demonstration of strong nonreciprocal thermal emission has been elusive, owing to the challenges of creating high-quality thin-film structures, and measuring emission over broad spectral and angular ranges while providing a substantial magnetic field. We note that nonreciprocal absorption was demonstrated at room temperature by measuring bi-directional reflectance [38-40], but emission of the sample was not measured.

In this Letter, we experimentally observe strong nonreciprocal thermal emission in a metamaterial using a custom-designed angle-resolved magnetic thermal emission spectroscopy (ARMTES). We achieve a difference between emissivity and absorptivity at the same wavelength and angle up to 0.43 on a gradient-doped epsilon-near-zero In$_{0.53}$Ga$_{0.47}$As (hereafter referred to as InGaAs) metamaterial atop gold. The metamaterial also shows significant nonreciprocal emission over broad spectral and angular ranges.

We designed and fabricated a gradient-doped epsilon-near-zero (ENZ) metamaterial, consisting of five 440-nm-thick electron-doped InGaAs layers atop 100-nm Au, as schematically shown in Fig. 1(b) with a scanning electron microscope cross-section. Details of sample fabrication are described in the Supplemental Material and Fig. S1 in [41]. The doping concentration increases as the depth increases, with concentrations denoted in Fig. 1(b). A thin film can support Berreman mode [48] at wavelengths where the permittivity is near zero. The gradient doping allows deeper layers to reach ENZ condition at progressively shorter wavelengths, creating cascaded Berreman modes. We transferred the epitaxially grown InGaAs layers to a Si substrate using thermocompression bonding [49,50], which allows for device integration for nonreciprocal-based applications.

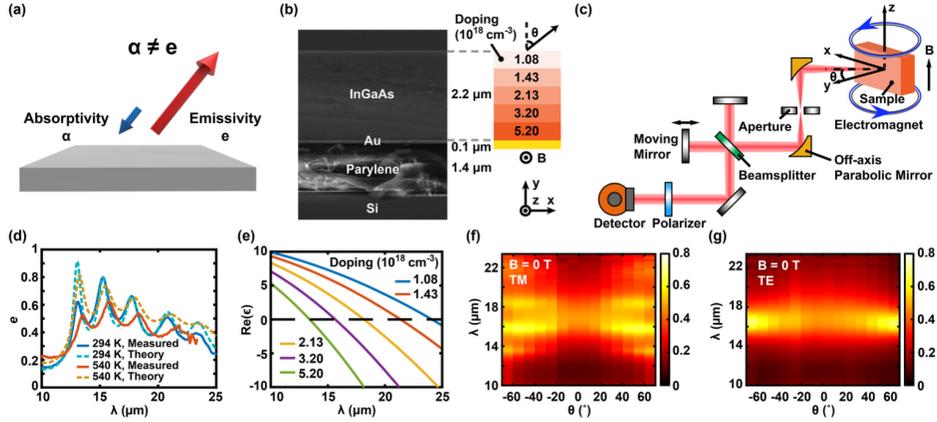

FIG. 1. Sample and experimental set-up. (a) Schematic of nonreciprocal emission and absorption. (b) Scanning electron microscope image of an InGaAs multilayer. It consists of five 440-nm-thick layers of gradient-doped InGaAs on top of 100-nm-thick gold. The doping concentration increases with the depth. (c) Schematic of ARMTES set-up. (d) Emissivities of the sample at TM polarization and 65°, measured at 540 K and obtained at room temperature (solid lines). Modeled emissivities at the corresponding temperatures are shown in dashed lines. (e) Spectra of real part of the permittivity $\epsilon$ of each InGaAs layer at 540 K. (f-g) Measured emissivities at 540 K, for (f) TM and (g) TE polarizations. In (d-g), $B = 0\,T$.

The permittivity of InGaAs in a magnetic field $\vec{B} = B\hat{z}$ is:

$$\bar{\bar{\epsilon}} = \epsilon_\infty \bar{\bar{I}} - \frac{\omega_p^2}{(\omega - i\Gamma)^2 - \omega_c^2} \begin{bmatrix} 1 - i\frac{\Gamma}{\omega} & i\frac{\omega_c}{\omega} & 0 \\ -i\frac{\omega_c}{\omega} & 1 - i\frac{\Gamma}{\omega} & 0 \\ 0 & 0 & \frac{(\omega - i\Gamma)^2 - \omega_c^2}{\omega(\omega - i\Gamma)} \end{bmatrix}.$$

Here, $\epsilon_\infty$ is the relative permittivity of undoped InGaAs, $\omega_p = \sqrt{\frac{nq^2}{m^*\epsilon_0}}$ is the plasma frequency, $\Gamma$ is the relaxation rate, $\omega_c = qB/m^*$ is the cyclotron frequency, $m^*$ is the electron effective mass, $q$ is the charge of a proton, and $n$ is the electron concentration. Details of permittivity model are described in [41]. In the magnetic field, the permittivity tensor becomes asymmetric, breaking Lorentz reciprocity. We use InGaAs because it has low electron effective mass [51] leading to high cyclotron frequency at a given magnetic field, and it can be epitaxially transferred for device integration [49].

To overcome the challenge of measuring thermal emission of a sample at various angles in a substantial magnetic field, we custom-designed an ARMTES set-up as depicted in Fig. 1(c). The set-up integrates a cryostat, an electromagnet, and an infrared spectrophotometer (see details of set-up in Appendix). We measure the spectral emissivity of the sample $e(\theta, B)$ for angle $\theta$ and magnetic field $B$, at 540 K (see details of measurement in Appendix). For a specular sample, thermodynamics requires that the emissivity equals the absorptivity at the opposite angle, leading to $e(\theta, -B) = \alpha(-\theta, -B)$ [26]. Further, due to the $C_2$ symmetry of the sample, after rotating both the angle and the magnetic field around the normal direction $y$ by 180°, the absorptivity must remain unchanged, leading to $\alpha(-\theta, -B) = \alpha(\theta, B)$. Using thermodynamics and rotational symmetry, we have $\alpha(\theta, B) = e(\theta, -B)$ for our sample [52,53]. Therefore, we determine absorptivity $\alpha(\theta, B)$, by measuring emissivity at the same angle but reversed magnetic field $e(\theta, -B)$. The nonreciprocity $[e(\theta, B) - \alpha(\theta, B)]$, is equivalent to the emissivity contrast $\Delta e = e(\theta, B) - e(\theta, -B)$.

We start by investigating emission at zero magnetic field. At 65° and TM polarization, a reflectance-based measurement at room temperature shows five emissivity peaks at wavelengths between 13 and 25 μm [Fig. 1(d)] (see details of measurement in [41] and Fig. S3 for all angles and polarizations). Each emissivity peak corresponds to the ENZ wavelength of an individual InGaAs layer [Fig. 1(e)]. Using ARMTES, at 540 K, we observe a redshift of the emissivity peaks [Fig. 1(d)]. The redshift can be understood largely by the increasing refractive index as the temperature rises [46] (see details in [41] and Fig. S4). For the topmost InGaAs layer, at room temperature, its ENZ wavelength corresponding to emissivity peak is at ~23.3 $\mu m$. At 540 $K$, the ENZ wavelength and accordingly the emissivity peak shifts to longer wavelength outside the spectral range of the detector. The measured emissivities agree well with modeling [dashed lines,

Fig. 1(d)] based on fluctuational electrodynamics [26,54]. The emissivity at 540 K without magnetic field exhibits mirror symmetry at both polarizations, that is $e(\theta, B) = e(-\theta, B)$ [Figs. 1(f-g)]. For the specular sample, as $e(-\theta, B) = \alpha(\theta, B)$ [26], the mirror symmetry of emissivity signifies reciprocal emission and absorption.

To break the reciprocity between emission and absorption, we apply $\pm 5\ T$ magnetic field to the sample at 540 K, and measure the emissivity. In a magnetic field, the emissivity in TM polarization becomes asymmetric [Figs. 2(a-b)]. At $5\ T$, there is strong broadband emission towards positive angles [Fig. 2(a)]. There are four emissivity peaks corresponding to Berreman mode of each InGaAs layer, covering a broad spectrum from 13 to 23 $\mu m$. In contrast, the emissivity is weak towards negative angles. However, at $-5\ T$, the emissivity shows opposite angular dependence [Fig. 2(b)]: the emissivity is high towards negative angles, but weak towards positive angles. Therefore, over a broad spectrum from 13 to 23 μm, there is strong contrast $\Delta e$ between the emissivities at $5\ T$ and $-5\ T$, which is equal to the nonreciprocity $[e(5T) - \alpha(5T)]$ [Fig. 2(c)]. Moreover, Fig. 2(d) shows that nonreciprocal emission persists over a broad angular range. At 21.6 μm and $5\ T$, nonreciprocity is 0.14 at 15°, reaches maximum of 0.43 at 55°, and remains >0.4 at 65°. Our experiment for the first time realizes strong nonreciprocal emission, with nonreciprocity as high as 0.43, which is much higher than nonreciprocity in literature [36,37]. Moreover, the achieved nonreciprocity with broad spectral (>10 μm) and angular ranges can provide strong nonreciprocal overall heat flux.

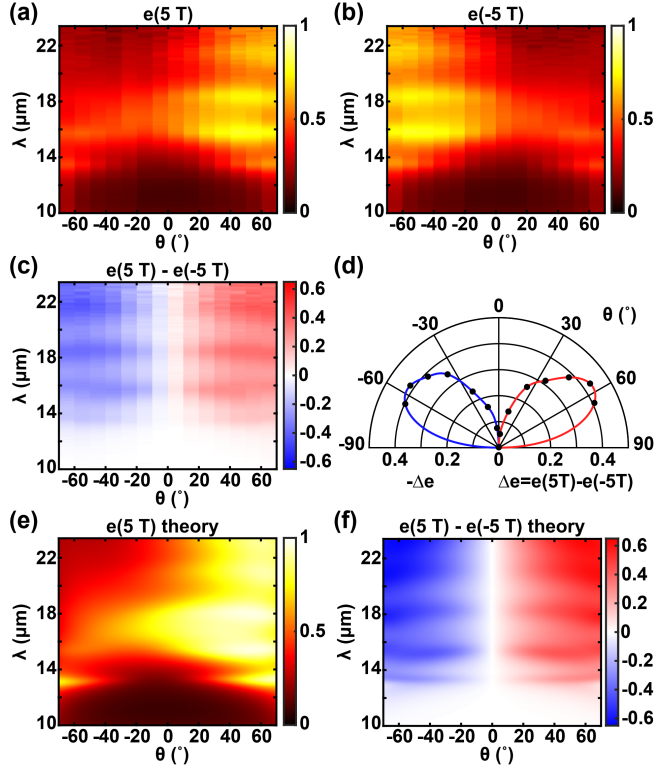

FIG. 2. Spectral and angular response of nonreciprocal emission. (a) Measured spectral, angular emissivity of the sample at 5 T magnetic field. (b) Measured spectral, angular emissivity at -5 T, which is equivalent to spectral, angular absorptivity at 5 T. (c) Difference between measured emissivity at 5 T and that at -5 T, which is equivalent to nonreciprocity. (d) Angular dependence of the contrast between the measured emissivity at 5 T and that at -5 T, at 21.6 μm. (e) Modeled emissivity at 5 T. (f) Difference between modeled emissivity at 5 T and that at -5 T. TM polarization is considered throughout (a-f).

To compare with the measurements, we model the emissivity using fluctuational electrodynamics [26,54]. Figures 2(e-f) show the modeled emissivity at $5\,T$, and the emissivity contrast between the cases of $5\,T$ and $-5\,T$. The modeling agree well with the measurements in Figs. 2(a) and 2(c), by considering increased relaxation rates at elevated temperature [44] (see permittivity model in [41]).

As nonreciprocal emission requires magnetic field, a key question is how the nonreciprocal emission depends on the magnetic field. To answer this we performed measurements at various magnetic fields from $-5\,T$ to $5\,T$ at TM polarization. Results of these measurements are shown in Figs. 3(a) and 3(d) for 55° and -55°, respectively (see Figs. S5-S16 in [41] for measurements at other angles). At 55°, as $B$ increases from $-5\,T$ to $5\,T$, the emissivity is enhanced over a broad spectrum, especially at the resonance peaks. The strong monotonic change of the emissivity as $B$ changes results in a substantial difference between emissivity at $B$ and that at $-B$ for 55°, which equals nonreciprocity [Fig. 3(b)]. The nonreciprocity is stronger at longer wavelengths, maximizing at the resonance wavelength corresponding to second topmost InGaAs layer. The nonreciprocity increases as $B$ increases. Figure 3(c) shows the measured emissivity contrast $\Delta e$ between $B$ and $-B$, at 55° and resonance wavelengths. For $B$ within 3 $T$, the nonreciprocity scales linearly as $B$. However, for $B$ larger than 3 $T$, the increase of the nonreciprocity becomes sublinear, especially at the two longer resonance wavelengths [Fig. 3(c)]. Such change of emissivity contrast from linear to sublinear with increasing $B$ is because emissivity contrast has an upper limit of 1. When the emissivity contrast becomes sufficiently high, the change in emissivity contrast needs to transition to sublinear.

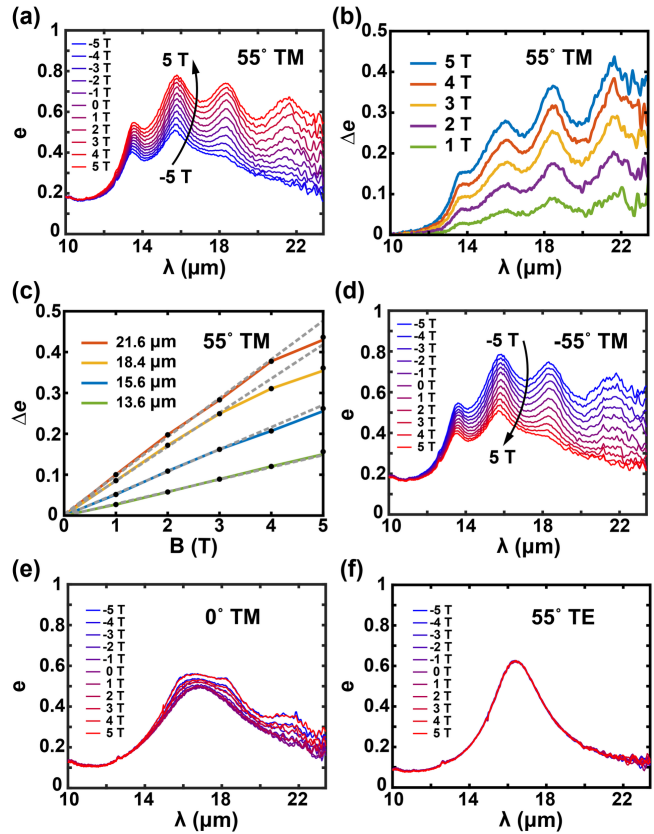

FIG. 3. Magnetic control of nonreciprocal emissivity. (a) Measured TM-polarized emissivity at 55° at varying magnetic fields. (b) The contrast between the measured emissivity at $B$ and the emissivity at $-B$, as a function of wavelength, for varying $B$ at 55° and TM polarization. (c) The magnetic dependence of emissivity contrast, at 55° and resonance wavelengths. The dashed lines denote linear fitting at $B \leq 3\ T$. (d-f) Measured emissivity for varying $B$, at (d) $-55°$ and TM polarization, (e) 0° and TM polarization, and (f) 55° and TE polarization.

The magnetic dependence of TM-polarized emissivity at -55° shows opposite trend compared to that at 55° [Fig. 3(d)]. At -55°, as $B$ varies from $-5\ T$ to $5\ T$, the TM-polarized emissivity decreases. At 0°, the TM-polarized emissivity slightly increases as the magnitude of magnetic field rises from $0\,T$ to $5\,T$ [Fig. 3(e)]. Such magnetic dependence is because at 0°, the TM-polarized emissivity is determined by $\epsilon_{xx} = \epsilon_\infty - \frac{\omega_p^2(\omega-i\Gamma)}{\omega[(\omega-i\Gamma)^2-\omega_c^2]}$, which slightly depends on $B$ through the cyclotron frequency $\omega_c$. In contrast to oblique angles, at 0°, the TM-polarized emissivity remains the same between $B$ and -$B$, which is consistent with the $C_2$ symmetry of the sample. The rotational symmetry requires the emissivity to remain unchanged by flipping both the angle and $B$, i.e., $e(\theta, B) = e(-\theta, -B)$, leading to $e(0°, B) = e(0°, -B)$ at $\theta = 0°$. Therefore, for our sample, achieving nonreciprocal emission in TM polarization requires both nonzero $B$ and angle.

We further investigate the magnetic dependence of TE-polarized emissivity. The TE-polarized emissivity at 55° remains unchanged as $B$ changes [Fig. 3(f)] (see Fig. S17 in [41] for measurement at 65°). The independence of TE-polarized emissivity from magnetic field is because TE-polarized emissivity is determined by $\epsilon_{zz}$ which is independent of $B$ along z-direction.

To gain further insights into nonreciprocal emission, and to understand the effects of magnetic field and angle, we examined detailed measurement results. Figures 4(a-d) show the measured emissivity in the $B$-$\theta$ plane at resonance wavelengths. The measured emissivity peaks at very positive $B$ and $\theta$, as well as at very negative B and $\theta$. The emissivity remains unchanged after flipping both $\theta$ and $B$, that is $e(\theta, B) = e(-\theta, -B)$, which is consistent with the $C_2$ symmetry of the sample.

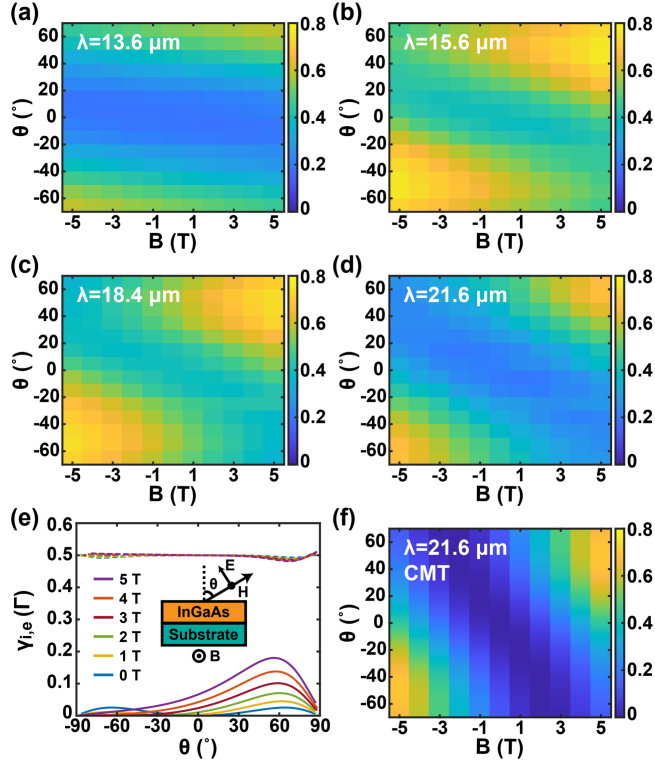

FIG. 4. Emissivity as a function of magnetic field and angle. (a-d) Measured emissivity in $B - \theta$ plane, at resonance wavelengths, including (a) 13.6 μm, (b) 15.6 μm, (c) 18.4 μm, and (d) 21.6 μm. (e) Intrinsic decay rate (dashed lines) and external decay rate (solid lines) for 440-nm-InGaAs atop a substrate at different angles and magnetic fields at TM polarization. The decay rates are normalized by the relaxation rate $\Gamma$. The calculation assumes that InGaAs has an electron concentration of $1.43 \times 10^{18}$ cm$^{-3}$, the substrate has a relative permittivity of -4, and the temperature is 540 K. (f) Modeled emissivity at 21.6 μm using coupled mode theory (CMT).

To understand the intricate dependence of emissivity on magnetic field and angle, we perform analysis using coupled mode theory (CMT) [55,56]. The peak emissivity of a resonant mode is $4\gamma_i\gamma_e/(\gamma_i + \gamma_e)^2$, where $\gamma_i$ and $\gamma_e$ are the intrinsic and the external modal decay rates, respectively. Unity emissivity is predicted when the intrinsic and the external decay rates match. For a single InGaAs layer, the decay rates at TM polarization are shown in Fig. 4(e). Details of calculation are described in [41]. While $\gamma_i$ remains about $\Gamma/2$ for all considered $\theta$ and $B$, $\gamma_e$ strongly varies with $\theta$ and $B$. With positive $B$ along $z$ direction, $\gamma_e$ at positive angle is higher than that at negative angle. We note that at 5 $T$, $\gamma_e$ is maximized at ~55°, leading to better matching of the two decay rates at 55° and a high emissivity. In contrast, due to the angular asymmetry of $\gamma_e$, the emissivity at -55° is weak. Using $e(-B, 55°) = e(B, -55°)$ owing to the $C_2$ symmetry, the coupled mode theory predicts that nonreciprocity which is equivalent to $e(B, \theta) - e(-B, \theta)$ should be maximized at ~55°, which agrees with Fig. 2(d). The emissivity in $B - \theta$ plane predicted from coupled mode theory for a thin film [Fig. 4(f)] shows good agreement with the measurement [Fig. 4(d)].

Using coupled mode theory, due to the approximate linear relationship of $\gamma_e$ with respect to $B$ and $\theta$ at positive magnetic field and $\theta \leq 60°$ [Fig. 4(e)], and the approximately constant $\gamma_i$, emissivity is determined by $B$ and $\theta$ through $\gamma_e$. Accordingly, at large positive $B$ and $\theta$, due to the high $\gamma_e$ and better matching between $\gamma_e$ and $\gamma_i$, there is high emissivity. The coupled mode theory further suggests that emissivity can be maintained by maintaining $\gamma_e$, through judiciously controlling $B$ and $\theta$. Indeed, we observe that the measured emissivity is largely maintained by adjusting $B$ and $\theta$ with slopes of $-1.5°/T$, $-5°/T$, $-6.5°/T$, and $-8°/T$ corresponding to Figs. 4(a-d), respectively. Therefore, magnetic field and angle can be tuned together to achieve novel control of emission.

We emphasize that the observed nonreciprocal thermal emission is due to Berreman modes. Surface modes and hyperbolic modes can be magnetically tuned to control near-field radiative heat transfer [57,58]. However, surface modes are not coupled to the far field in planar sample, and hyperbolic modes are not supported in our experiment where wavevector is perpendicular to the magnetic field [57]. Therefore, surface and hyperbolic modes are not responsible for the measured emission. Also, while we measured nonreciprocal emission from directions that are perpendicular to the in-plane magnetic field, emission in a plane that is parallel to magnetic field has been theoretically shown to be reciprocal [59]. We also note that nonreciprocal circularly polarized emission has been predicted in the Faraday configuration where the magnetic field is perpendicular to the surface of magneto-optical materials [24].

In summary, we demonstrated strong nonreciprocal thermal emission on a gradient epsilon-near-zero InGaAs multilayer. Significant nonreciprocal emission also persists over broad spectral and angular ranges. Different from all prior studies on nonreciprocal emitters and absorbers [36-40], we transferred epitaxial structure to a foreign substrate, enabling its device integration for nonreciprocal-based applications. Due to the use of metal reflector enabled by epitaxial transfer, at wavelengths where emission is largely reciprocal such as 8.5-13 μm, our sample shows negligible parasitic reciprocal emission (Fig. S3), which can be useful for applications such as nonreciprocal heat flux control. We note that a magneto-optical film atop a spacer and substrate has been predicted to support strong nonreciprocal emission and absorption [60], but over a narrow spectrum. Also, while InAs and InSb could be used to design nonreciprocal emitters, epitaxial transfer is not straightforward for InAs and InSb. We also note that strong nonreciprocal emission has been predicted with smaller magnetic field [27] which can be generated by permanent magnets, or by using magnetic Weyl semimetals even without magnetic field [33,34]. The role of thermal nonreciprocity in radiative cooling has been previously discussed in Refs. [61-63], with related perspectives and developments presented in these works. The demonstrated strong nonreciprocal emission over broad spectral and angular ranges can be useful for creating nonreciprocal-based applications such as heat flux control, energy conversion, and sensing [64].

**Acknowledgments**

L.Z. acknowledge support from National Science Foundation award no. 2238927, the Kaufman New Investigator Award through the Charles E. Kaufman Foundation - a supporting organization of the Pittsburgh Foundation, and Penn State Institute of Energy and the Environment (IEE) Seed Grant. We thank Dr. Bangzhi Liu in Penn State Materials Research Institute for support in scanning electron microscopy. We acknowledge use of Penn State Nanofabrication Lab and Materials Characterization Lab for fabricating and characterizing the sample.

**End Matter**

*Appendix - Measurement using the angle-resolved magnetic thermal emission spectroscopy:* The angle-resolved magnetic thermal emission spectroscopy system integrates a cryostat (Advanced Research Systems), a superconducting electromagnet (Cryomagnetics), and a Fourier-transform infrared spectrophotometer (Bruker Invenio X). The sample and a high-emissivity coating (Acktar Metal Velvet) are mounted on the front and the back sides of a temperature-controlled, fully-rotatable sample holder using ceramic adhesive (PELCO). The sample holder is placed in a high vacuum ($<1 \times 10^{-6}\ Torr$) in the cryostat and the sample holder can be rotated 360° without breaking the vacuum. In experiment, the extended portion of the cryostat containing with the sample holder is inserted to a bore of the electromagnet. The electromagnet provides magnetic field up to $\pm 5\ T$, allowing for fully exploring magnetic control of nonreciprocal emission. The temperature of the sample controlled using a temperature controller (Lakeshore 336). The thermal radiation of the sample propagates through an orthogonal bore of the electromagnet, is reflected off two off-axis parabolic mirrors, through an interferometer and an infrared linear polarizer, and collected by a mercury-cadmium telluride (MCT) detector, as shown in Fig. 1(c). To eliminate the possibility of collecting thermal radiation of the sample that is reflected by the inner surface wall of the cryostat, the cryostat's inner surface wall is coated with high-emissivity coating due to its near-unity absorptivity (see [41] and Fig. S2). The cryostat is equipped with diamond window, which allows for performing measurement at the whole spectral range of the MCT detector (D315).

For measuring the emissivity of the sample, we first measure the radiation signal $I_S$ from the sample heated to temperature $T_S = 540\ K$. Then we perform a reference measurement by measuring the radiation signal $I_R$ of the high-emissivity coating that is mounted on the sample holder at $T_S$. Finally, to measure background signal $I_{BG}$, we measure the radiation from high-emissivity coating placed in ambient air (~294 K). The spectral, angular emissivity of the sample $e$ is determined as:

$$e(\lambda, \theta, B, T_s) = e_R(\lambda) \cdot \frac{I_S(\lambda, \theta, B, T_s) - I_{BG}(\lambda, 294\ K)}{I_R(\lambda, T_s) - I_{BG}(\lambda, 294\ K)},$$

where $e_R$ is the emissivity of the high-emissivity coating. Details of the emissivity and measurement of the coating are described in [41] and Fig. S2.